\documentclass[preprint]{aastex}

\usepackage{graphicx}
\usepackage{epstopdf}

\newcommand{\kms}{\ensuremath{\rm km\,s^{-1}}}
\newcommand{\ms}{\ensuremath{\rm m\,s^{-1}}}
\newcommand{\gcmc}{\ensuremath{\rm g\,cm^{-3}}}

\newcommand{\teff}{\ensuremath{T_{\rm eff}}}
\newcommand{\logg}{\ensuremath{\log{g}}}
\newcommand{\vsini}{\ensuremath{v \sin{i}}}
\newcommand{\feh}{[Fe/H]}

\newcommand{\rsun}{\ensuremath{R_\sun}}
\newcommand{\msun}{\ensuremath{M_\sun}}
\newcommand{\lsun}{\ensuremath{L_\sun}}

\newcommand{\rstar}{\ensuremath{R_\star}}
\newcommand{\mstar}{\ensuremath{M_\star}}
\newcommand{\loggstar}{\ensuremath{\logg_\star}}
\newcommand{\lstar}{\ensuremath{L_\star}}

\newcommand{\rpl}{\ensuremath{R_{\rm P}}}
\newcommand{\mpl}{\ensuremath{M_{\rm P}}}

\newcommand{\rhopl}{\ensuremath{\rho_{\rm P}}}
\newcommand{\loggpl}{\ensuremath{\logg_{\rm P}}}
\newcommand{\teq}{\ensuremath{T_{\rm eq}}}

\newcommand{\rjup}{\ensuremath{R_{\rm J}}}
\newcommand{\mjup}{\ensuremath{M_{\rm J}}}

\newcommand{\koicur}{Kepler-8}
\newcommand{\koicurb}{Kepler-8b}

\newcommand{\koicurCCra}{\ensuremath{18^{\mathrm{h}}45^{\mathrm{m}}09^{\mathrm{s}}.15}}
\newcommand{\koicurCCdec}{\ensuremath{+42^{\circ}27'03''.9}}
\newcommand{\koicurCCkic}{KIC~6922244}
\newcommand{\koicurCCtwomass}{2MASS~18450914+4227038}
\newcommand{\koicurCCkicr}{13.511} 
\newcommand{\koicurCCkicg}{13.89} 

\newcommand{\koicurLCar}{\ensuremath{6.97^{+0.20}_{-0.24}}}			
\newcommand{\koicurLCrprstar}{\ensuremath{0.09809^{+0.00040}_{-0.00046}}}	
\newcommand{\koicurLCimp}{\ensuremath{0.724\pm{0.020}}}   			
\newcommand{\koicurLCi}{\ensuremath{84.07\pm{+0.33}}}			

\newcommand{\koicurLCP}{\ensuremath{3.52254^{+0.00003}_{-0.00005}}}		
\newcommand{\koicurLCPshort}{3.523}						
\newcommand{\koicurLCPprec}{\ensuremath{3.52254}}				%
\newcommand{\koicurLCT}{\ensuremath{2454954.1182^{+0.0003}_{-0.0004}}}		
\newcommand{\koicurLCdip}{\ensuremath{9.82\pm0.22}}		

\newcommand{\koicurSMEteff}{\ensuremath{6213\pm150}}	
\newcommand{\koicurSMEfeh}{\ensuremath{-0.055\pm0.03}}	
\newcommand{\koicurSMElogg}{\ensuremath{4.28\pm0.10}}	
\newcommand{\koicurSMEvsini}{\ensuremath{10.5\pm0.7}}	

\newcommand{\koicurYYm}{\ensuremath{1.213^{+0.067}_{-0.063}}}		
\newcommand{\koicurYYmshort}{\ensuremath{1.213}}			
\newcommand{\koicurYYmlong}{\ensuremath{1.213^{+0.067}_{-0.063}}}	
\newcommand{\koicurYYrlong}{\ensuremath{1.486^{+0.053}_{-0.062}}}	
\newcommand{\koicurYYlogg}{\ensuremath{4.174\pm0.026}}			
\newcommand{\koicurYYlum}{\ensuremath{4.03^{+0.52}_{-0.54}}}		
\newcommand{\koicurYYmv}{\ensuremath{3.28 \pm 0.15}}                 
\newcommand{\koicurYYage}{\ensuremath{3.84\pm1.5}}			
\newcommand{\koicurXdist}{\ensuremath{1330\pm 180}}             

\newcommand{\koicurRVK}{\ensuremath{68.4\pm12.0}}                     
\newcommand{\koicurRVmean}{\ensuremath{-52.72\pm0.10}}       

\newcommand{\koicurPPlogg}{\ensuremath{2.871\pm0.119}}			
\newcommand{\koicurPParel}{\ensuremath{0.0483^{+0.0006}_{-0.0012}}}	
\newcommand{\koicurPPrho}{\ensuremath{0.261\pm0.071}}                  
\newcommand{\koicurPPrhoshort}{\ensuremath{0.26}}			

\newcommand{\koicurPPm}{\ensuremath{0.603^{+0.13}_{-0.19}}}		
\newcommand{\koicurPPmshort}{\ensuremath{0.60}}				
\newcommand{\koicurPPmlong}{\ensuremath{0.603^{+0.13}_{-0.19}}}		
\newcommand{\koicurPPr}{\ensuremath{1.419^{+0.056}_{-0.058}}}		
\newcommand{\koicurPPrshort}{\ensuremath{1.419}}			
\newcommand{\koicurPPrlong}{\ensuremath{1.419^{+0.056}_{-0.058}}}	
\newcommand{\koicurPPteq}{\ensuremath{1764 \pm 200}}              

\shortauthors{Jenkins et al.}
\shorttitle{\koicurb}

\begin{document}

\title{Discovery and Rossiter-McLaughlin Effect \\ 
of Exoplanet \koicurb \altaffilmark{0}}

\altaffiltext{0}{Based in part on observations obtained at the W.~M.~Keck Observatory,
which is operated as a scientific partnership among the California Institute of
Technology, the University of California, and the National Aeronautics
and Space Administration. The Observatory was made possible by the
generous financial support of the W.M. Keck Foundation.}

\author{
Jon~M.~Jenkins\altaffilmark{2},
William~J.~Borucki\altaffilmark{1},
David~G.~Koch\altaffilmark{1},
Geoffrey~W.~Marcy\altaffilmark{3},
William~D.~Cochran\altaffilmark{8},
Gibor Basri,\altaffilmark{3}
Natalie~M.~Batalha\altaffilmark{4},
Lars~A.~Buchhave\altaffilmark{5,6},
Tim~M.~Brown\altaffilmark{7},
Douglas~A.~Caldwell\altaffilmark{2},
Edward~W.~Dunham\altaffilmark{9},
Michael~Endl\altaffilmark{8},
Debra~A.~Fischer\altaffilmark{10},
Thomas~N.~Gautier III\altaffilmark{11},
John~C.~Geary\altaffilmark{5},
Ronald~L.~Gilliland\altaffilmark{12},
Steve~B.~Howell\altaffilmark{13},
Howard~Isaacson\altaffilmark{3},
John~Asher~Johnson\altaffilmark{14},
David~W.~Latham\altaffilmark{5},
Jack~J.~Lissauer\altaffilmark{1},
David~G.~Monet\altaffilmark{15},
Jason~F.~Rowe\altaffilmark{1,16},
Dimitar~D.~Sasselov\altaffilmark{5},
William~F.Welsh\altaffilmark{17},
Andrew~W.~Howard\altaffilmark{3},
Phillip~MacQueen\altaffilmark{8},
Hema~Chandrasekaran\altaffilmark{2}, 
Joseph~D.~Twicken\altaffilmark{2}, 
Stephen~T.~Bryson\altaffilmark{1}, 
Elisa~V.~Quintana\altaffilmark{2}, 
Bruce~D.~Clarke\altaffilmark{2}, 
Jie~Li\altaffilmark{2}, 
Christopher~Allen\altaffilmark{18}, 
Peter~Tenenbaum\altaffilmark{2}, 
Hayley~Wu\altaffilmark{2}, 
Soren~Meibom\altaffilmark{5},
Todd~C.~Klaus\altaffilmark{18}, 
Christopher~K.~Middour\altaffilmark{18}, 
Miles~T.~Cote\altaffilmark{1}, 
Sean~McCauliff\altaffilmark{18}, 
Forrest~R.~Girouard\altaffilmark{18}, 
Jay~P.~Gunter\altaffilmark{18}, 
Bill~Wohler\altaffilmark{18}, 
Jennifer~R.~Hall\altaffilmark{18}, 
Khadeejah~Ibrahim\altaffilmark{18}, 
AKM~Kamal~Uddin\altaffilmark{18}, 
Michael~S.~Wu\altaffilmark{19}, 
Paresh~A.~Bhavsar\altaffilmark{1}, 
Jeffrey~Van~Cleve\altaffilmark{2}, 
David~L.~Pletcher\altaffilmark{1}, 
Jessie~A.~Dotson\altaffilmark{1}, 
Michael~R.~Haas\altaffilmark{1}}
\altaffiltext{1}{NASA Ames Research Center, Moffett Field, CA 94035}
\altaffiltext{2}{SETI Institute/NASA Ames Research Center, Moffett Field, CA 94035}
\altaffiltext{3}{University of California, Berkeley, Berkeley, CA 94720}
\altaffiltext{4}{San Jose State University, San Jose, CA 95192}
\altaffiltext{5}{Harvard-Smithsonian Center for Astrophysics,
60 Garden Street, Cambridge, MA 02138}
\altaffiltext{6}{Niels Bohr Institute, Copenhagen University, DK-2100 Copenhagen, Denmark}
\altaffiltext{7}{Las Cumbres Observatory Global Telescope, Goleta, CA 93117}
\altaffiltext{8}{University of Texas, Austin, TX 78712}
\altaffiltext{9}{Lowell Observatory, Flagstaff, AZ 86001}
\altaffiltext{10}{Radcliffe Institute, Cambridge, MA \& Yale
  University, New Haven, CT}
\altaffiltext{11}{Jet Propulsion Laboratory/California Institute of Technology, Pasadena, CA 91109}
\altaffiltext{12}{Space Telescope Science Institute, Baltimore, MD 21218}
\altaffiltext{13}{National Optical Astronomy Observatory, Tucson, AZ 85719}
\altaffiltext{14}{California Institute of Technology, Pasadena, CA 91109}
\altaffiltext{15}{US Naval Observatory, Flagstaff Station, Flagstaff, AZ 86001}
\altaffiltext{16}{NASA Postdoctoral Fellow Program}
\altaffiltext{17}{San Diego State University, San Diego, CA}
\altaffiltext{18}{Orbital Sciences Corporation/NASA Ames Research Center, M/S 244-30, Moffett Field, CA 94035, USA}
\altaffiltext{19}{Bastion Technologies/NASA Ames Research Center, M/S 244-30, Moffett Field, CA 94035, USA}


\begin{abstract}
We report the discovery and the Rossiter-McLaughlin effect of
\koicurb, a transiting planet identified by the NASA \emph{Kepler Mission}.
\emph{Kepler} photometry and Keck-HIRES radial velocities yield the radius
and mass of the planet around this F8IV subgiant host star.  The
planet has a radius $\rpl = \koicurPPrshort\,\rjup$ and a mass, $\mpl
= \koicurPPmshort\,\mjup$, yielding a density of \koicurPPrhoshort\,\gcmc, among the lowest density planets known.
The orbital period is $P = \koicurLCPshort$ days and orbital semimajor
axis is \koicurPParel AU.  The star has a large rotational \vsini \ of
\koicurSMEvsini \ \kms and is relatively faint (V $\approx$
\koicurCCkicg \ mag), both properties deleterious to precise Doppler
measurements.  The velocities are indeed noisy, with scatter of 30 \ms,
but exhibit a period and phase consistent with the planet implied by
the photometry.  We securely detect the Rossiter-McLaughlin effect,
confirming the planet's existence and establishing its orbit as
prograde.  We measure an inclination between the projected planetary
orbital axis and the projected stellar rotation axis of $\lambda =
-26.9 \pm 4.6^{\circ}$, indicating a moderate inclination of the
planetary orbit.  Rossiter-McLaughlin measurements of a large sample of
transiting planets from \emph{Kepler} will provide a statistically robust
measure of the true distribution of spin-orbit orientations for hot
jupiters in general.
\end{abstract}

\section{Introduction}

To date, 90 ``hot jupiters''---gas giant planets with periods
$\leq$10~days---have been detected around Sun-like stars
\citep{torres08}.  The front-running formation scenario supposes that
these planets did not form where they reside today, close the host
star, because the inner regions of protoplanetary disks have
inadequate surface densities and high temperatures \citep{lin96}.
Instead hot jupiters are presumed to form several astronomical units
(AU) from their host stars followed by subsequent migration inward to
their current locations.  One likely migration scenario involves tidal
interactions between the planet and a remaining gaseous disk
\citep{lin96,moorhead08}, causing the planet to spiral inward while
maintaining its nearly circular orbit that is co-planar with the disk.
Alternatively, migration may occur by N-body gravitational
interactions, such as planet--planet scattering
\citep{rasio96,chat07}, dynamical relaxation \citep{pap01,adams03},
and Kozai interactions with a distant object, and damped later by
tidal friction \citep{holman97,fab07,wu07,nagasawa08}.  Thus,
measurements of both the orbital eccentricities and the orbital
inclinations relative to the star's equator offer diagnostics of the
original migration process.

We assume that planets form in protoplanetary disks with the stellar
spin and planetary orbital axes aligned.  If so, the nearly adiabatic tidal
interactions between planets and disks would maintain the alignment
\citep{ward94}.  In contrast, few-body gravitational interactions
would typically cause misalignments.  Few-body models by
\citet{adams03} predict a final inclination distribution for
dynamically relaxed planetary systems that peaks near $20^{\circ}$ and
which extends to inclinations as high as $85^{\circ}$.  Kozai
interactions between a planet and an outer body (star or planet) result
in a wide distribution of final orbital inclinations for the inner
planet, including retrograde orbits \citep{fab07, wu07, nagasawa08}.

The Rossiter-McLaughlin (R-M) effect offers a way to assess
quantitatively the spin-orbit alignment of a planetary system by
measuring the Doppler effect of the star's light during a planetary
transit. As the planet blocks receding portion of a rotating star's
surface, the spectrum from the unobscured surface has a net Doppler
shift toward shorter wavelengths, and vice versa for blocking the
approaching portion of the star. The R-M effect has been measured in
18 stars to date \citep{queloz00, winn05, winn06b, winn07a, 
wolf07, narita07, narita08, bouchy08, Cochran08, lo08,winn08, Johnson09,
Pont09a, Moutou09, Pont09a, Winn09b, narita09a, narita09b,
Simpson10, Anderson:09, Gillon09, Pont09b, Amaury10}.

About 2/3 of the 18 planetary systems measured by the R-M effect have
an orbital plane well aligned with the star's equatorial plane, as
projected onto the sky, giving $\lambda$ near $0^\circ$. This
alignment is as expected from simple migration theory due to gentle
loss of orbital energy to the gas in the protoplnaetary disk
\citep{lin96}.  However, six exoplanetary systems show a significant
spin-orbit misalignment, namely HD 80606 \citep{Winn09a, Moutou09,
  Pont09a, Gillon09}, WASP-14b \citep{Johnson09, Joshi09}, XO-3b
\citep{Hebrard09, Winn09}, HAT-P-7b \citep{Winn09b, narita09a}, CoRoT-1
\citep{Pont09b} Wasp-17b \citep{Anderson:09}.

The variety of alignments support the bimodal distribution found by
\citet{Fabrycky09}.  Interestingly, all four misaligned systems
contain quite massive planets, above 1 \mjup .  This correlation may
be related to the association of massive planets with higher orbital
eccentricity \citep{Wright09}, as both eccentricity and inclination
may arise from perturbations of planets from their original circular
orbits.  But there are two massive planets on eccentric orbits for
which $\lambda$ appears to be consistent with zero, namely HD 17156b
\citep{Cochran08, Barbieri09, narita09b}, and HATp--2b
\citep{winn07,lo08}.  There is currently no dominant and secure
explanation for the misaligned or eccentric hot jupiters.

Here, we present the first detection of the Rossiter-McLaughlin effect
from a planet detected by the \emph{Kepler Mission}.  As this mission is
expected to detect dozens of transiting hot jupiters, \emph{Kepler} offers
an opportunity to provide a statistically robust measure
of the distribution of spin-orbit angles, and to correlate that
angle with other physical properties of the systems.

\section{\emph{Kepler} Photometry}

Nearly continuous photometry in a 100 square degrees field near Cygnus and
Lyra was carried out during 42 days by the \emph{Kepler} spaceborne
telescope, as described previously \citep{Borucki:10, Borucki_kepler4,
  Koch_kepler5, Jenkins:10, Batalha:10, Gautier:10}.  The star \koicur
\ (= \koicurCCkic, $\alpha = \koicurCCra, \delta = \koicurCCdec$,
J2000, KIC $r = \koicurCCkicr$\,mag) exhibits a repeated dimming of
\koicurLCdip millimag, obvious against uncertainties in each 30 minute
integration of 0.1 millimag.  The light curve for \koicur \ is plotted in Figure
\ref{fig:lightcurve}.  The numerical data are available electronically
in the online edition of the journal.  A modest amount of detrending has been applied
\citep{Koch:10, Rowe:10} to the time series.

We detect no systematic difference between alternating transit events
at 50 $\mu$mag levels, ruling out nearly equal components of an
eclipsing binary star, (see Fig.\,\ref{fig:lightcurve}).  We also see
no evidence of dimming at the expected times of a secondary eclipse,
which would be visible for most eclipsing binary systems of unequal
surface brightness.  The photocenter shows no displacement
astrometrically above millipixel levels (0.5 mas) during times in and
out of transit as would be seen if there were a background eclipsing
binary masquerading as a transiting planet.  Thus the photometric and
astrometric non-detections of an eclipsing binary support the planet
interpretation for the repeated transit signatures.  Moreover, the
shape of the photometric transit is adequately fit with a
planet-transit model further supporting the planet interpretation.

We fit the light curves by solving for a/R$_{\rm Star}$, the density
of the star, and the ratio of planet to stellar radius.  We follow the
method for estimating stellar radii and other stellar parameters
described by \citet{soz07}, \citet{bak07}, \citet{winn07c} and
\citet{cha07}.  This method extracts physical properties directly from
the light curve, geometry, and Newtonian physics, and it uses the
greater orthogonality of parameters to yield more robust fits to
observables \citep{torres08}.  We fit explicitly for $a/$\rstar, $b$,
and $r/$\rstar using a procedure developed by one of us (J.Rowe) and
described by \citet{Koch:10, Borucki_kepler4}.

We begin with an LTE spectroscopic analysis \citep{Valenti:96,
  Valenti:05} of a high resolution template spectrum from Keck-HIRES of
Kepler-8 to derive an effective temperature, \teff = \koicurSMEteff
\ K, surface gravity, \logg = \koicurSMElogg (cgs), metallicity, \ and
the associated error distribution for each of them.  The multitude of
Yale-Yonsei stellar evolution models \citep{Demarque04} are
constrained by both those LTE measurements and by the stellar density
that stems directly from the orbital period, the fractional dimming
during transit, and measures of transit durations \citep{soz07,
  Brown:10}.  By Monte Carlo analysis, those photometric and
spectroscopic constraints and their uncertainties establish the probabiliby
density contours among the evolutionary tracks where the star may
reside.  We iterate the self-consistent fitting of light curves,
radial velocities, and evolutionary models until a domain of stellar
mass, radius, and age is identified.  That domain encompasses a range
of evolutionary states that satisfy all of the constraints within
their error distributions.  The resulting mass and radius of \koicur
\ are given in Table 1, along with other associated stellar properties
such as luminosity and age.

\section{A Background  Eclipsing Binary: Follow-up Imaging}

We carried out extensive tests of the possibility that the apparent
photometric transit was actually caused by a background eclipsing
binary star within the photometric aperture of radius $\sim$8 \arcsec.
We obtained images with 0.8 arcsec seeing with the Keck telescope
HIRES guider camera and the bg38 filter to search for stellar
companions that might be eclipsing.  This filter combined with the CCD
detector have a response similar to the V plus R bands, in turn
similar to the bandpass of the \emph{Kepler} photometer.  This Keck image is shown in
\ref{fig:keck_guider}.  There is one star having 0.0075 the flux of
the main star (V and R band) that resides 3.8\arcsec northwest of \koicur.
This background star resides within the \emph{Kepler} aperture and could
conceiveably be an eclipsing binary that is masquerading as a
transiting planet.  

But this faint neighboring star cannot be the source of the
photometric transit signature for two reasons.  If the background star
were the cause of the observed 1\% dimming, the photocenter centroid
shifts would be 3-20 millipixel on the \emph{Kepler} CCD.  Instead,
astrometric measurements in and out of transit \citep{Jenkins:10} show
shifts of no more than 0.1 millipixel in and out of transit.  Figure
\ref{fig:CentroidTimeSeries} shows both the flux and astrometric
photocenter (centroid) of the \emph{Kepler} images during quarter 1 month of
data.  We applied a high-pass filter to remove non-transit signatures
on timescales longer than 2 days.  At times of transits there are no
displacements in either the row or column direction at a level above a
millipixel.  This indicates that any background eclipsing binary would
need to be well within $\sim$0.1 pixels or 0.4\arcsec of the
target star in order to explain the photometric transit signals.

To hunt further for background eclipsing binaries we plotted flux
versus the photocenter centroids in both the row and column
directions, as shown in Figure \ref{fig:RainPlot}.   These so-called
``rain plots'' would reveal an eclipsing binary as a ``breeze'' in the
centroids to the left or right as the flux drops.   No such breeze is
detected at a level near 0.1 millipixel, ruling out all but eclipsing
binaries located within a few tenths of an arcsec of \koicur.  
Finally, we looked directly at the \emph{Kepler} images taken both during and out of
transit to detect motion of the centroid of light, as shown in Figure \ref{fig:DiffImage}.
We formed the difference of the images in and out of transit to detect
astrometric displacements associated with the flux dimming, as would
occur if a neighboring eclipsing binary were the cause.
Those difference images show no shift of the photocenter.   We conclude that the
transit photometry with its 1\% dimming cannot be explained by any
eclisping binary companions beyond 1 arcsecond of \koicur.

Furthermore, the flux ratio of the two stars is only 0.0075, making it
impossible for the background star to cause the 1\% photometric
dimming.  Even if that background star were to vanish, the total flux
would decline by less than the observed 1\%.  We conclude that the 1\%
photometric dimming is not caused by an eclipsing binary star within
the \emph{Kepler} photometric aperture, from 1 - 10 arcsec of the target
star.

To hunt for additional stars in the field within an arcsec of \koicur \ 
we used both speckle imaging and adaptive optics imaging.  \koicur \ was
observed at the Palomar Hale 200-inch telescope on 09 Sep 2009 UT with Palomar
near-infrared adaptive optics system [PHARO] \citep{Hayward:01}. The
PHARO instrument was utilized in the J-band filter with the 25 mas per
pixel (25\arcsec FOV) mode.  The source was observed with a 5-point
dither and an integration time of 2.8 seconds per frame.  The dither
pattern was repeated 7 times for a total on-source integration time of
98 seconds.  The average uncorrected seeing during the observations at
J-band was $0.65\arcsec$, and the average AO corrected images produced
point spread functions that were 0.09\arcsec FWHM.  There are two
faint sources within 4\arcsec of Kepler-8; these objects are 7 and 8.4
magnitudes fainter than the primary target and are too faint to
produce the \emph{Kepler}-observed transit and centroid shift.  The AO
imaging detected no sources at J-band down to within
$0.1\arcsec-0.2\arcsec$ of the primary target that are within $\Delta
m \approx 6-7$ mag at J.

Speckle observations of \koicur \ were made on 1 October 2009 (UT) at
the WIYN observatory located on Kitt Peak.  The observations were made
using the WIYN speckle camera during a night of very good seeing (0.62
\arcsec) and under clear conditions.  We used a narrow bandpass 40 nm
wide centered at 692 nm. The Kepler speckle program obtains
observations  of both double and single standard stars throughout the
night.  We use a robust background estimator on the reconstructed
images to set a limit for the level of companion star we should detect
if present. The speckle observations show that \koicur \ has no
companion star between 0.05 and 2.0 arcsec within a delta magnitude of
$<$4.2 mags.  These AO and speckle observations effectively rule out the
possibility of an eclipsing binary star to within as close as 0.1
arcsec from \koicur.

\subsection{Covariance of Inclination, Limb Darkening, and Stellar Parameters}

As \koicur \ has a large impact parameter, the solution is quite
sensitive to errors in limb-darkening.  The derived impact parameter
and inclination angle are directly related to the assumed
limb-darkening law, poorly known for the wide \emph{Kepler} bandpass.  We
adopted an adhoc approach to estimate the limb-darkening parameters
\citep{Rowe:10} as follows.  We noted that fits to the observed
Rossiter-McLaughlin effect (see section 6), coupled with the measured
\vsini = \koicurSMEvsini \ \kms offered a constraint on the
inclination angle and the impact parameter, and hence on the inferred
stellar radius, for a given limb-darkening law.  To solve the genereal
case for eccentric orbits see \citet{Pal10}.

For the cases of Kepler-4 and Kepler-5 \citep{Borucki_kepler4,
  Koch_kepler5}, we discovered that a model using tabular values \citep{Prsa06} for
limb darkening over-predicted the amount of curvature in the variation
of flux as a function of time during transit.  We therefore modified
the limb-darkening parameters by fitting the three known exoplanets,
TRES-2, HAT-P-7, and HAT-P-11, and using published values for their
stellar and planetary parameters.  We linearly interpolated over the
values of effective temperature to derive a superior measure of the
true limb-darkening law for the \emph{Kepler} bandpass. The photometry from
new transiting planets found by \emph{Kepler}, along with IR photometry that
is less sensitive to limb darkening, will allow us to refine the
limb-darkening law for the \emph{Kepler} bandpass in the coming months.  The
resulting fits to the photometric transits of Kepler-4, 5, 6, 7, and 8
were all improved with our new limb-darkening treatment
\citep{Rowe:10}.

For \koicur, the resulting impact parameter and inclination were
$b$=\koicurLCimp \ and $i$ = \koicurLCi \ both about 10\% smaller than
we had obtained by using the first-guess limb-darkening law.  The transit
duration in turn dictates the best-fit stellar radius, and after
iterating with the Yale-Yonsei models, the resulting stellar radius is
$R$ = \koicurYYrlong \ \rsun.  The stellar mass has a value within a narrow
range, $M$ = \koicurYYmlong \ \msun, again derived from iteration between the
fit to the light curve and to the Yale-Yonsei models.  Additional
information on stellar mass comes from the projected rapid rotation of
the star, \vsini = \koicurSMEvsini \ \kms .  This \vsini is faster
than typical rotation for stars at the lower end of the mass range
quoted above, especially as some evolution and associated increase in
moment of inertia has occurred.  Thus we marginally favor the upper
half of the quoted mass range, i.e. $M >$ \koicurYYmshort \ \msun.  The
stellar gravity is \logg = \koicurYYlogg (cgs), and the approximate age is
\koicurYYage Gyr, both indicating a star nearing the end of its main
sequence lifetime.

\section{Radial Velocities}

We took high resolution spectra of \koicur \ using HIRES on the Keck~I
10-m telescope \citep{vogt94}. We set up the HIRES spectrometer in the
same manner that has been used consistently for 10 years with the
California planet search \citep{Marcy08}.  We employed the red
cross-disperser and used the I$_2$ absorption cell to measure the
instrumental profile and the wavelength scale. The slit width was set
at $0\farcs87$ by the ``B5 decker'', giving a resolving power of about 60,000
at 5500\AA\.  Between 2009 June 1 and 2009 October 31 we gathered
15 spectra of \koicur \ out of transit (See  Figure\,\ref{fig:rv_vs_time}).  Typical
exposure times were between 10 - 45 min, yielding signal-to-noise of
20 - 40 ~pixel$^{-1}$.  These are quite low S/N ratios with which to
attempt precise Doppler measurements.  We tested the Doppler precision
from such low SNR spectra on standard stars, notably HD 182488 and HD
9407, and found that the expected errors are $\sim$5 \ms, as expected
from Poisson statistics of the photons, with no systematic errors
above 1 \ms, as usual for the iodine technique.   On 2009 October 29 (UT) we
obtained nine spectra of \koicur \ during transit (the last taken
during egress). The observations were taken between airmass 1.9 and
3.8, while the star was setting, the last exposure occurring at hour
angle 5 hr 35 min with extreme atmospheric refraction and dispersion.
Again, tests with standard stars observed at such high airmass show no systematic errors in the
Doppler measurements above 1 \ms.

We carried out careful reduction of the raw images, including cosmic ray
elimination and optimal extraction of the spectra, to minimize the
background moonlight.  We performed the Doppler analysis with the
algorithm of \citet{Johnson09}.  We estimated the measurement error in
the Doppler shift derived from a given spectrum based on the weighted
standard deviation of the mean among the solutions for individual
2~\AA\ spectral segments. The typical internal measurement error was
7-10~m~s$^{-1}$.  The resulting velocities are given in Table 2 and
plotted in Figure\,\ref{fig:rv_vs_time}.

The actual uncertainties in the velocities are certainly closer to
20-25\,\ms \ for several reasons.  The spectra have such low photon
levels, $\sim$200 photons per pixel in the raw CCD images, that cosmic
rays and background sky are significant noise sources.  The latter was
especially problematic as we used an entrance slit with dimensions
0.87 x 3.5 arcsec, not long enough to separate the wings of the star's
point spread function from the background sky.  Nearly all HIRES
spectra were taken with the moon gibbous or full, and about half of
the nights had moderate cirrus that scatters moonlight into
the slit.  A few observations made with a 14 arcsec slit revealed that
1-3\% of the light came from the moonlit sky in typical observations.
Simulations with stellar spectra contaminated by moonlit sky suggested
that errors of $\sim$10 \ms would accrue, no doubt systematic as well.
Errors in the velocities are large because of relatively rapid
rotation, \vsini = \koicurSMEvsini, broadening the absorption lines by
5x that of the slowest rotating stars.  Doppler errors increase
linearly with line widths.  Thus, while the S/N=40 might be expected
to yield velocities of $\sim$5 \ms, the high \vsini \ increases that
error to $\sim$25 \ms.  Indeed, the best orbital fit to the velocities,
shown in Figure\,\ref{fig:rv_vs_time}, exhibit discrepancies of
20-30\,\ms.  We estimate that the true errors in velocities are thus
$\sim$25\,\ms.  We have accounted for these errors by adding a
"jitter" of 25\,\ms to the internal velocity errors.  Those augmented
uncertainties are reflected in Table 2.

We carried out a Levenberg-Marquardt least-squares fit of a Keplerian,
single-planet model to the observed velocities.  In all models the
orbital period and the time of mid-transit was constrained to be that
found in the photometric fit.  We first assumed a circular orbit leaving
only two free parameters in our fit, namely the velocity amplitude and
the gamma velocity of the system.  The best-fit model is overplotted
in Figure\,\ref{fig:rv_vs_time}.  The best-fit value of the amplitude
is $K = \koicurRVK$\,\ms.  This, coupled with the adopted stellar mass
of \koicurYYm \ \msun, yields a planet mass of \koicurPPm \mjup.

We also carried out fits in which the eccentricity was allowed to
float.  The best-fit model has an eccentricity = 0.24, $K = 71.4$ \ms,
yielding a planet mass of 0.62 \mjup.  The RMS of the residuals is 39
\ms compared with 40 \ms \ for the circular orbit, rendering the
eccentric orbit no better than the circular.  Indeed a bootstrap Monte
Carlo estimate of the parameter uncertainties gives a formal error in
eccentricity of 0.16 .  We conclude that the eccentricity is
consistent with zero, but could be as high as 0.4.  This large
uncertainty stems from the high \vsini \ of the star and it's relative
faintness, V = 13.9 mag.

A line bisector analysis showed no variation at a level of a few meters
per second and no correlation with measured radial velocities.  Thus,
the velocity variation appears to represent actual acceleration of the
center of mass of the star.

\section{Properties of the Planet Kepler-8b}

With the planet interpretation highly likely, the properties of the
host star and the depth of transit directly yield a planet radius of
\koicurPPrlong $\rjup$ and a mass of \koicurPPmlong $\mjup$ .  The
errors represent the 68\%-probability domain of integrated uncertainty
from all input measurements and models.  The planet density is
\koicurPPrho \ \gcmc apparently placing it among the low density exoplanets.
Assuming a bond albedo of 0.1, its orbital radius of \koicurPParel
\ AU, and the best-fit stellar luminosity from the Yale-Yonsei models
gives an equilibrium temperature for the planet of \koicurPPteq \ K,
assuming rapid and complete redistribution of thermal energy around
the planet's surface.  This planet apparently is a member of the
population of low density, bloated hot jupiters having thermal
histories yet to be firmly understood.  The planet properties are
listed in Table 1.

\section{Rossiter-McLaughlin Effect\label{sec:rossiter-mclaughlin}}
The radial velocities obtained during the transit of 2009 October 29 (UT) 
were modeled using the techniques described by \citet{Cochran08}.
The models adopted the system parameters in Table\,\ref{tab:parameters},
and searched for the value of $\lambda$ that best matched the observed
velocities.   In particular, we assumed the zero eccentricity orbit, as
the \emph{Kepler} photometric lightcurve gives no indication of non-zero
eccentricity.  We used a \citet{claret04} four parameter limb-darkening law,
although running similar models with both a linear and a quadratic
limb-darkening law gave essentially the same results.
We derive a value of $\lambda = -26.9 \pm 4.6^{\circ}$ for the angle
between the projected stellar rotation axis and the projected planetary
orbital axis. The model fit to the data gives a reduced chi-squared
$\chi^2_0 = 0.66$ for the velocities obtained during the transit,
indicating that the uncertainties in these velocities are probably
slightly overestimated. 

The radial velocities measured during the transit
and the Rossiter-McLaughlin effect model fit are shown in
Figure\,\ref{fig:rossiter}.  The observed asymmetric Rossiter-McLaughlin
effect, with a positive deviation of larger amplitude and longer duration
than the negative amplitude is a result of the combination of the large
impact parameter and the non-zero value of $\lambda$.   A central transit
would give a symmetric Rossiter-McLaughlin deviation, with the amplitude
of the effect then depending on $\lambda$.  However, for an off-center transit, 
the overall shape depends critically on both the impact parameter and
on $\lambda$ (cf. Figure~2 of \citet{gaudi07}). 

This Rossiter-McLaughlin analysis provides independent verification of
the major results of the transit lightcurve analysis.  The lightcurve analysis
alone suggested a transit chord that is significantly off-center, with impact parameter,
$b = \koicurLCimp$ and inclination, $i = \koicurLCi$.  This large
impact parameter coupled with the non-zero value of $\lambda$ results
in the planet transit blocking mostly the approaching (blue-shifted)
half of the rotating star, as indicated in \ref{fig:transit_sketch}.
The planet crosses to the receding
(red-shifted) portion of the stellar disk just before the end of the
transit.  This causes the observed asymmetry in the Rossiter
McLaughlin velocity perturbation during the transit.  The observed
amplitude of the R-M effect is in excellent agreement with the
photometrically determined stellar radius and impact paramter, and
with the spectroscopically determined \vsini.  The duration of the
photometric transit agrees with the duration of the observed R-M
velocity perturbation.  This consistency between the properties of the
system as derived from the transit photometry and from the R-M
velocity perturbation gives a confirmation that both phenomena were
caused by an orbiting, planet-sized companion to the star. There is no other
explanation of the observed R-M variations that is also consistent with
both the photometric and the radial velocity observations.

This observation of the Rossiter-McLaughlin effect in \koicur \ is
important because it confirms that the observed lightcurve
and radial velocity variations in \koicur \ were indeed caused by an
orbiting planetary companion.  The amount of scatter in the radial
velocity data giving the solution shown in Figure~\ref{fig:rv_vs_time}
is uncomfortably large, even given the lack of line bisector
variations in these spectra.  However, the R-M data absolutely confirm
the existence of the planet.  There is no other explanation of the
observed R-M variations that is also consistent with both the
photometric, astrometric, imaging, and the radial velocity observations.  

\section{DISCUSSION\label{sec:discussion}}

\begin{figure}
\plotone{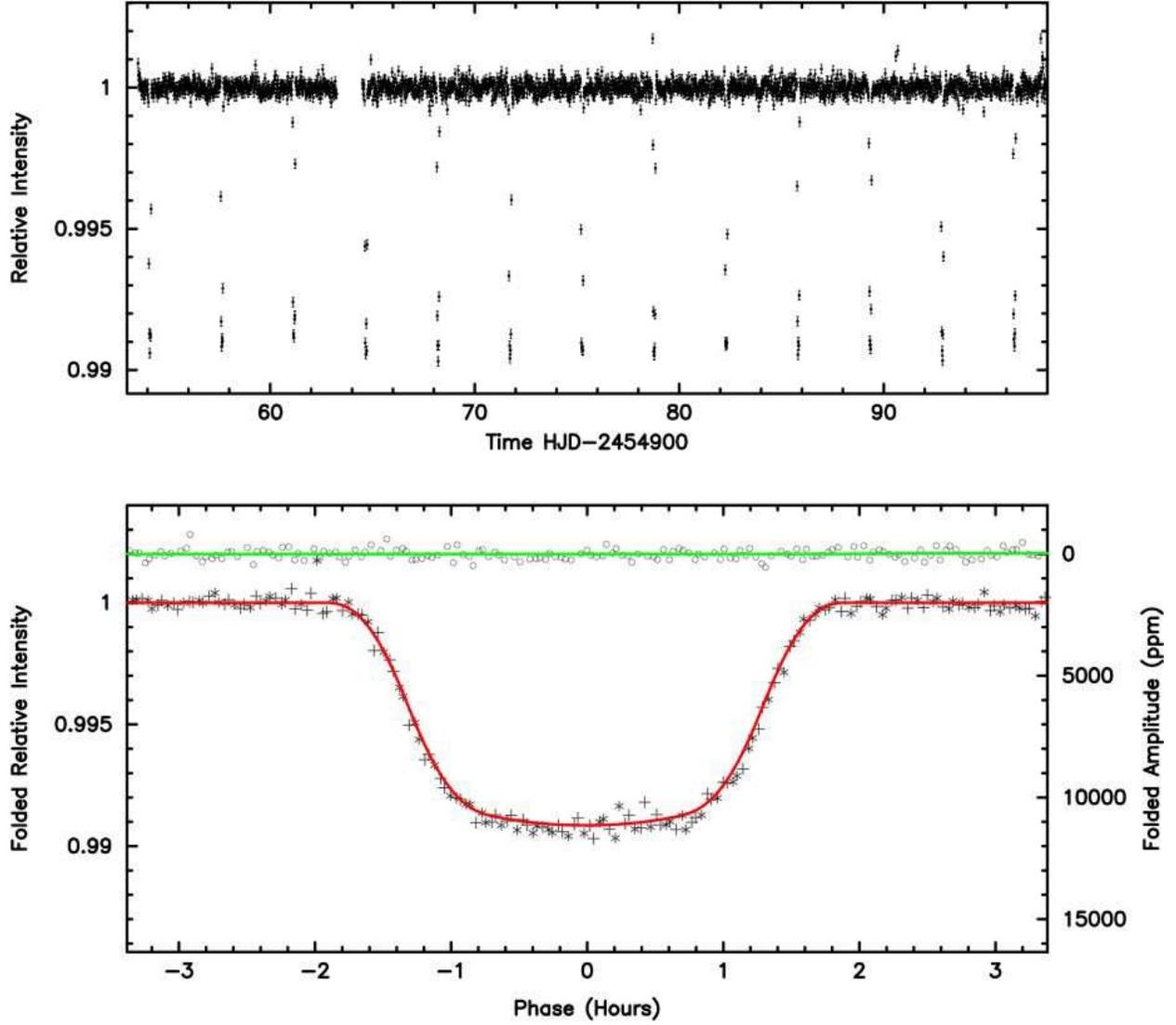}
\caption{
The detrended light curve for \koicur.  The time series for the entire
\emph{Kepler} photometric data set is plotted in the upper panel.  The lower panel shows the
photometry folded by the period $P = \koicurLCPprec$\,days, and the model
fit to the primary transit is plotted as a solid line.  Data for even and for odd transits are denoted by '+'s and '*'s, respectively. The even and odd transits do not exhibit significant variations in transit depth indicative of the primary and secondary eclipses of an eclipsing binary. 
The photometry at the predicted time of occultation of the planet is
shown just above the phased transit curve, showing no significant dimming at the level of 0.1 mmag.
The lack of any dimming at the predicted time of occulation, along
with the absence of both astrometric motion in and out of transit and
the absence of detected stellar companions, leaves the planetary interpretation
as the only plausible model, confirmed by the Rossiter-McLaughlin detection.
\label{fig:lightcurve}}
\end{figure}

\begin{figure}
\plotone{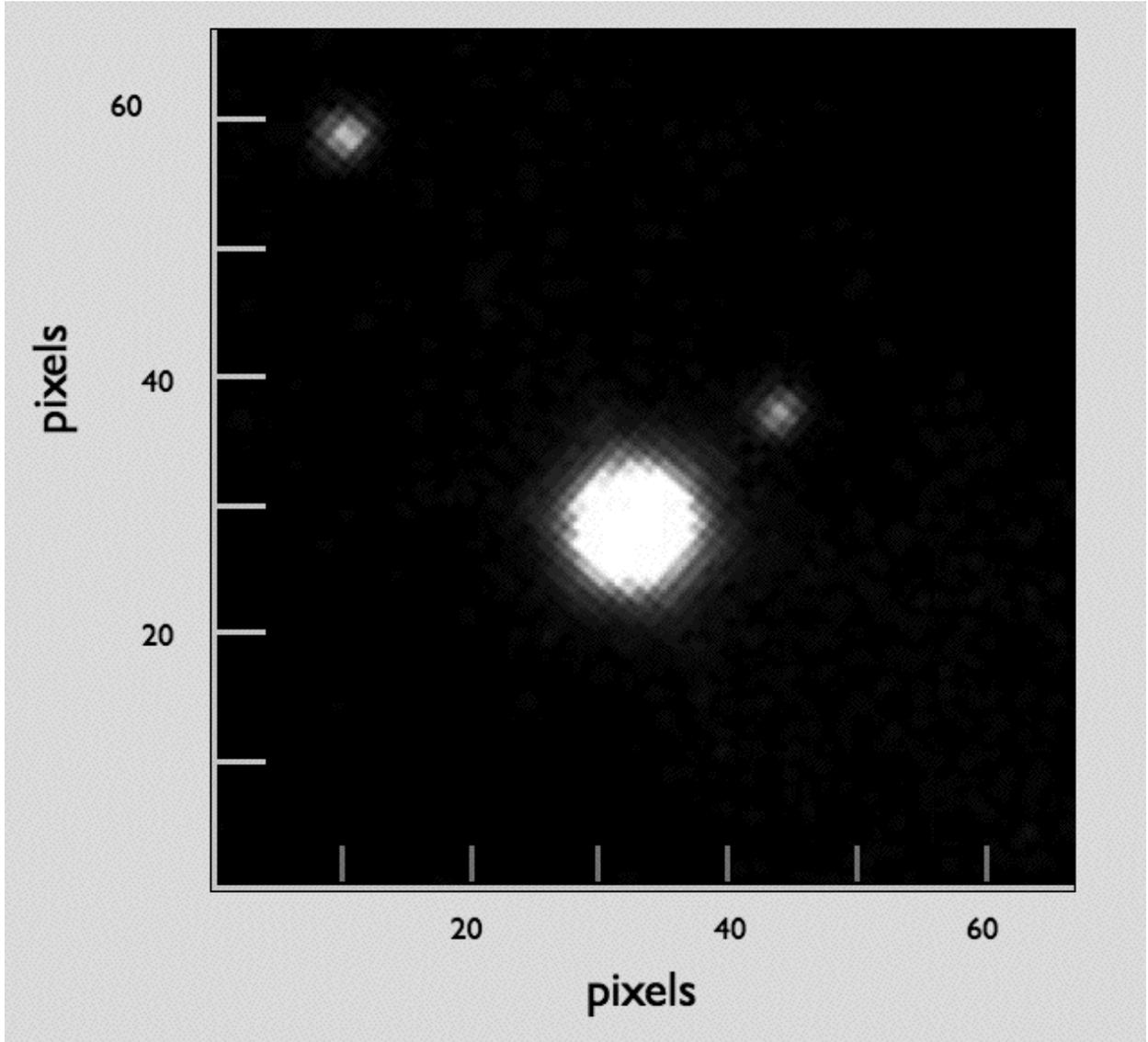}
\caption{
The image of the field near \koicur \ obtained from the Keck 1
telescope guider camera with a bg38 filter, making the bandpass
approximately V and R combined, similar to that of the \emph{Kepler}
photometer.  North is up, and east to the left, and the plate scale is
0.397 \arcsec per pixel.  The main star, \koicur \ is the brightest
source at the center of the image.  A faint star is located 3.8
\arcsec to the northwest of \koicur, having a flux 0.75\% that of
\koicur \ and residing within the \emph{Kepler} photometric aperture.  With a
flux less than 1\% that of \koicur, the companion cannot be
responsible for the periodici 1\% dimming, even if it were a
background eclipsing binary.  But this neighboring star is
sufficiently bright to account for the astrometric displacement of the
\koicur \ that occurs during transit of its planet, explaining the shift
of the photocenter of the two stars by 0.0001 \emph{Kepler} pixels.
\label{fig:keck_guider}}
\end{figure}

\begin{figure}
\plotone{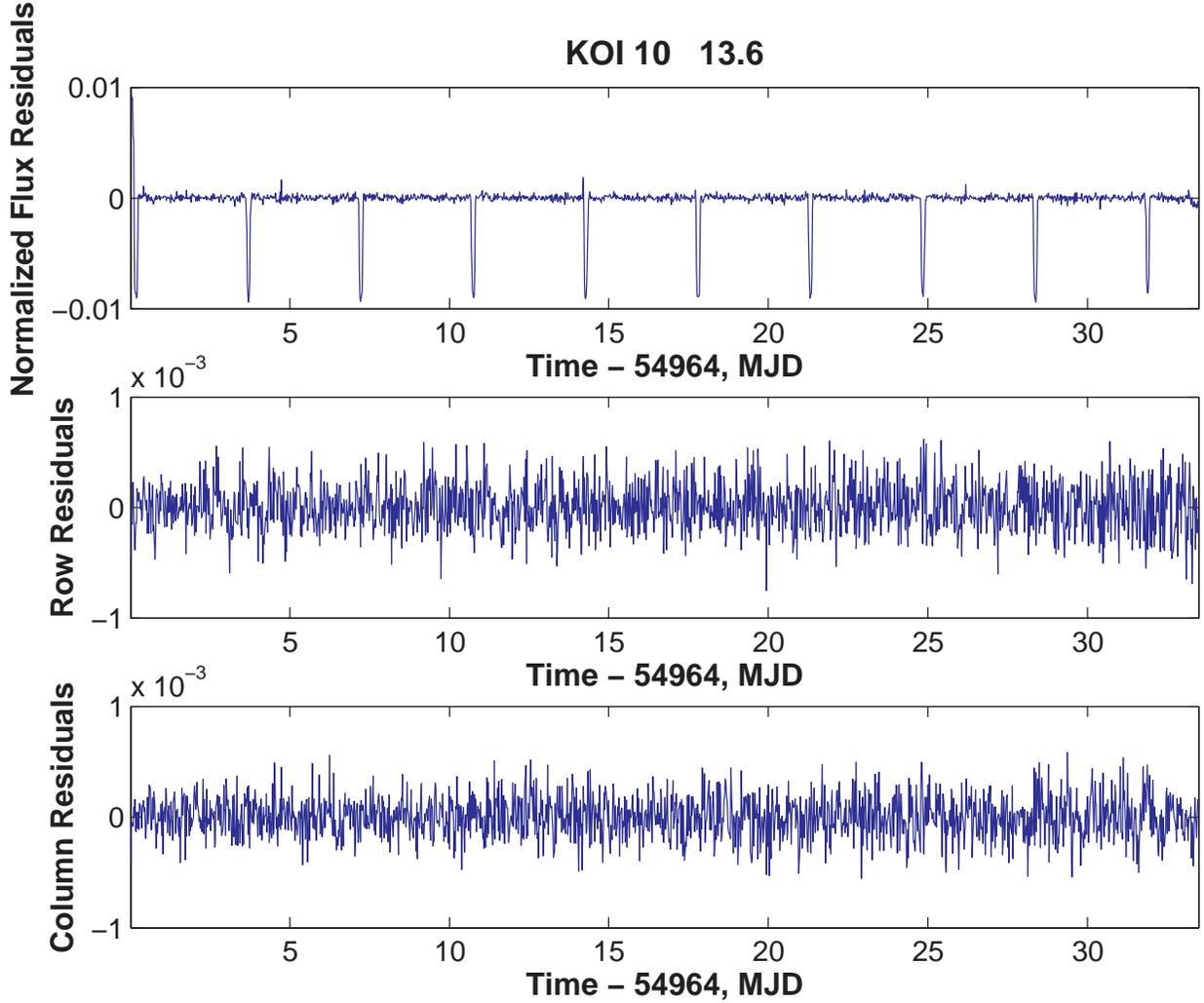}
\caption{Flux(upper) and astrometric centroid (middle, lower) timeseries of
  Kepler-8b from \emph{Kepler} quarter-1 data. A high-pass filter has been applied to each time series
  to remove non-transit signatures on timescales longer than 2
  days. There is no obvious displacement of the centroid of light in either the row
  or column residuals.  This lack of motion indicates that any background
  eclipsing binary would need to be located within $\sim$0.001/.01 = 0.1
  pixels corresponding to 0.4 \arcsec of the target star in order to
  explain the photometric transit signals but avoid showing motion.
\label{fig:CentroidTimeSeries}}
\end{figure}

\begin{figure}
\plotone{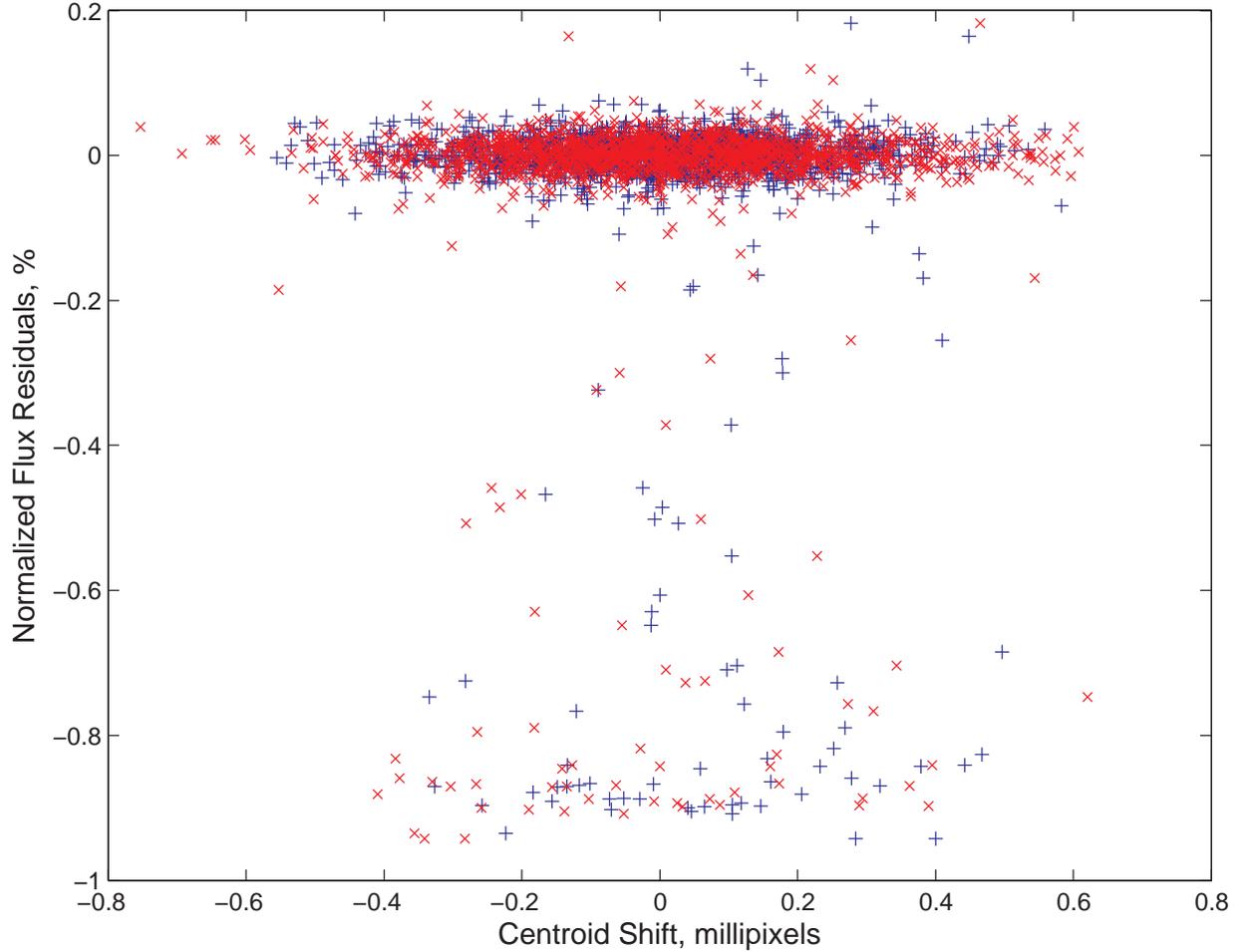}
\caption{Flux versus residual astrometric centroids in row (x) and
  column (+) for \koicur. The measurements are the same as those presented in Figure \ref{fig:CentroidTimeSeries}. This ``rainplot'' shows no evidence that the  centroids are systematically shifting during each transit event, which would be indicative of a blend with a background star. If this were the case, identification of the actual star responsible for the transit features in the light curve would require inspection of a high resolution image or catalog to unravel the mystery. Since the in-transit points "rain" straight down, rather than slanting along a diagonal, any blend scenario would require a very small separation on the sky between the two sources.
\label{fig:RainPlot}}
\end{figure}

\begin{figure}
\plotone{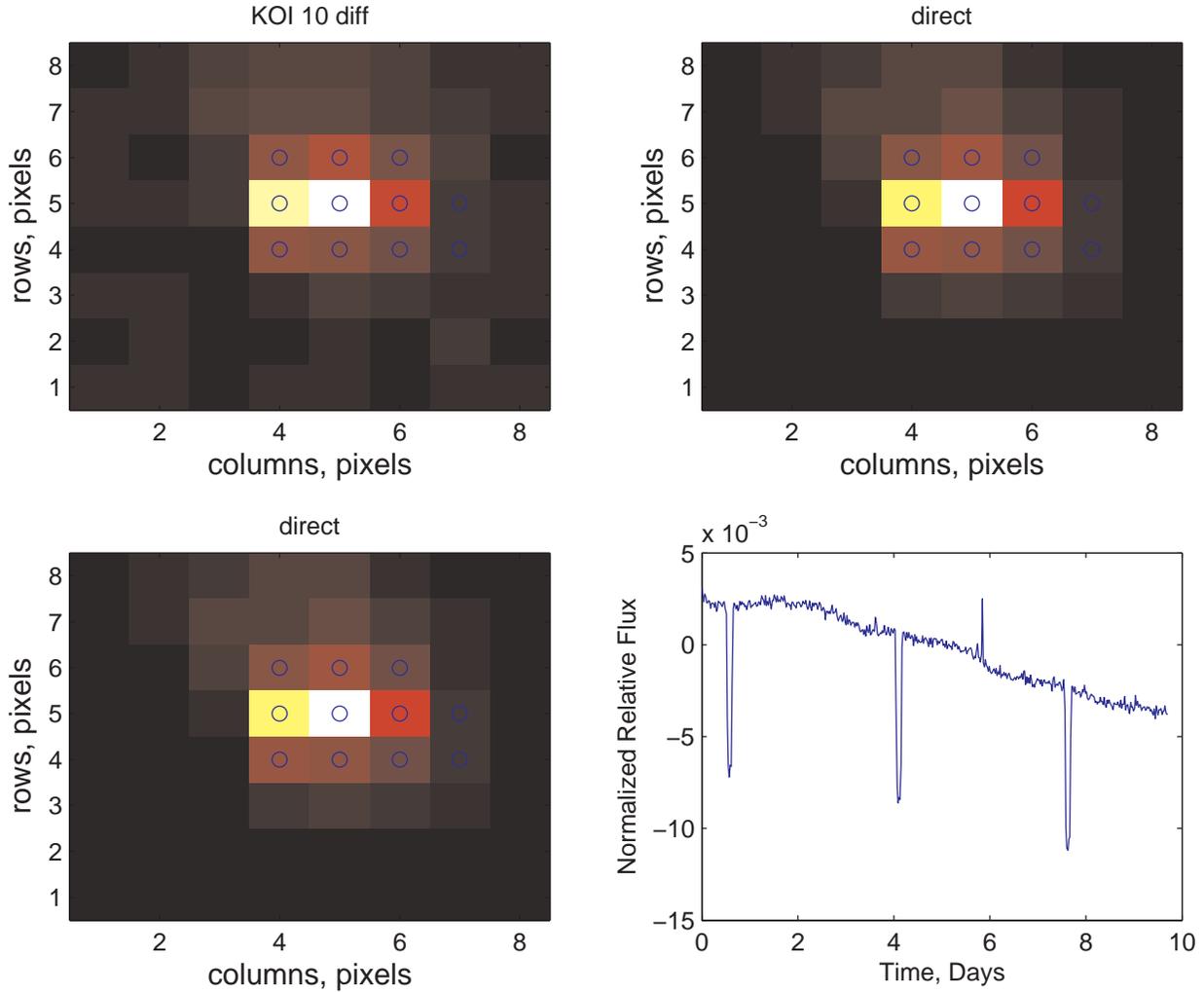}
\caption{The results of a "poor man's" difference image analysis are presented in this figure: (A) (upper left) is the result of subtracting  the average of the in-transit images from the average of a similar number of just out-of-transit images for one of Kepler-8b's transits. The pixels summed to form the photometric brightness measurements are indicated by the blue circles. For comparison, the average of the out-of-transit images are displayed in (B -- upper right) and (C -- lower left). The raw flux light curve from Q0 is displayed panel (D -- lower right). The difference image, Panel (A) is remarkably similar to the direct images. Had the transits been due to a background eclipsing binary that were 0.5 or more pixels offset from the target star, then the difference image would display a noticeable shift in the location of the pixel containing the peak energy. Instead, both difference and direct images supporting a coincident localization of the source of the transits to the target star, Kepler-8.
\label{fig:DiffImage}}
\end{figure}

\begin{figure}
\plotone{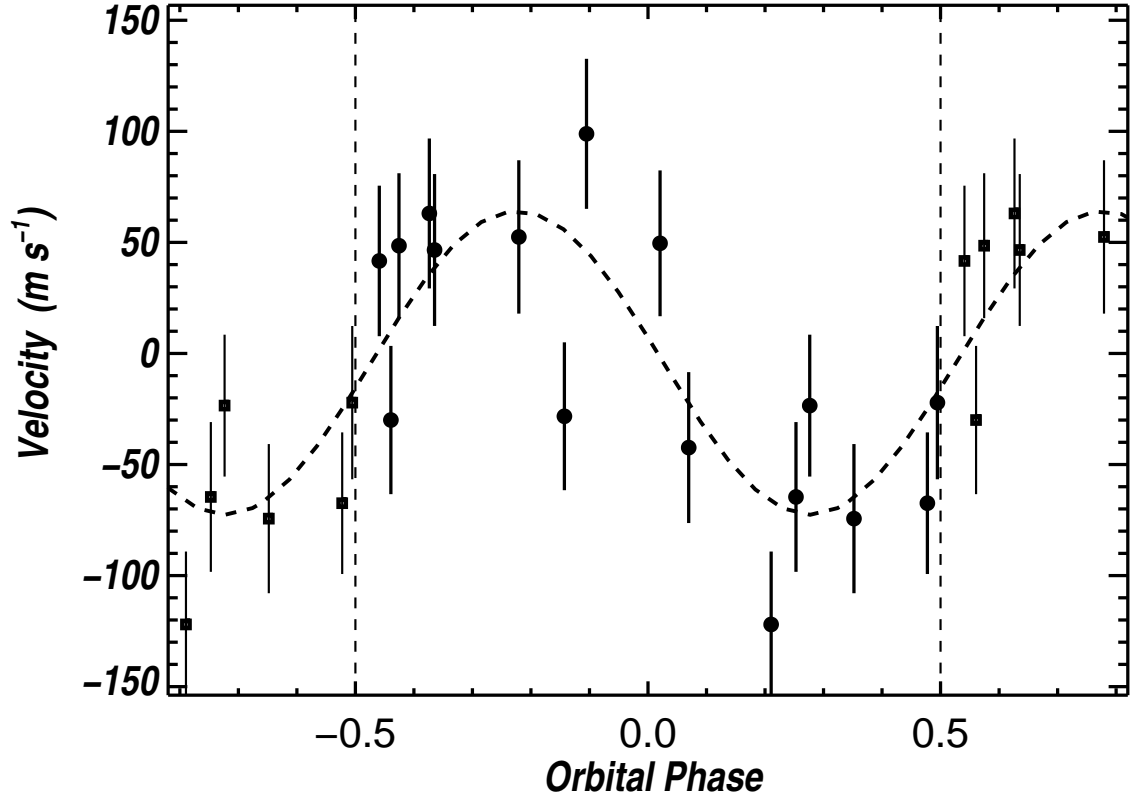}
\caption{
The orbital solution for \koicur. The observed radial velocities
obtained with HIRES on the Keck 1 Telescope are plotted
together with the best-fit velocity curve for a circular orbit with the period
and time of transit fixed by the photometric ephemeris.  Uncertainties
are large due to the faintness of the star (V=13.9 mag), the high
rotational Doppler broadening (\vsini = \koicurSMEvsini ), and
contamination from background moonlight in most spectra.  The measured velocities vary
high and low as predicted by the orbit from the photometric transit
curve, supporting the planet model.
\label{fig:rv_vs_time}}
\end{figure}

\begin{figure}
\plotone{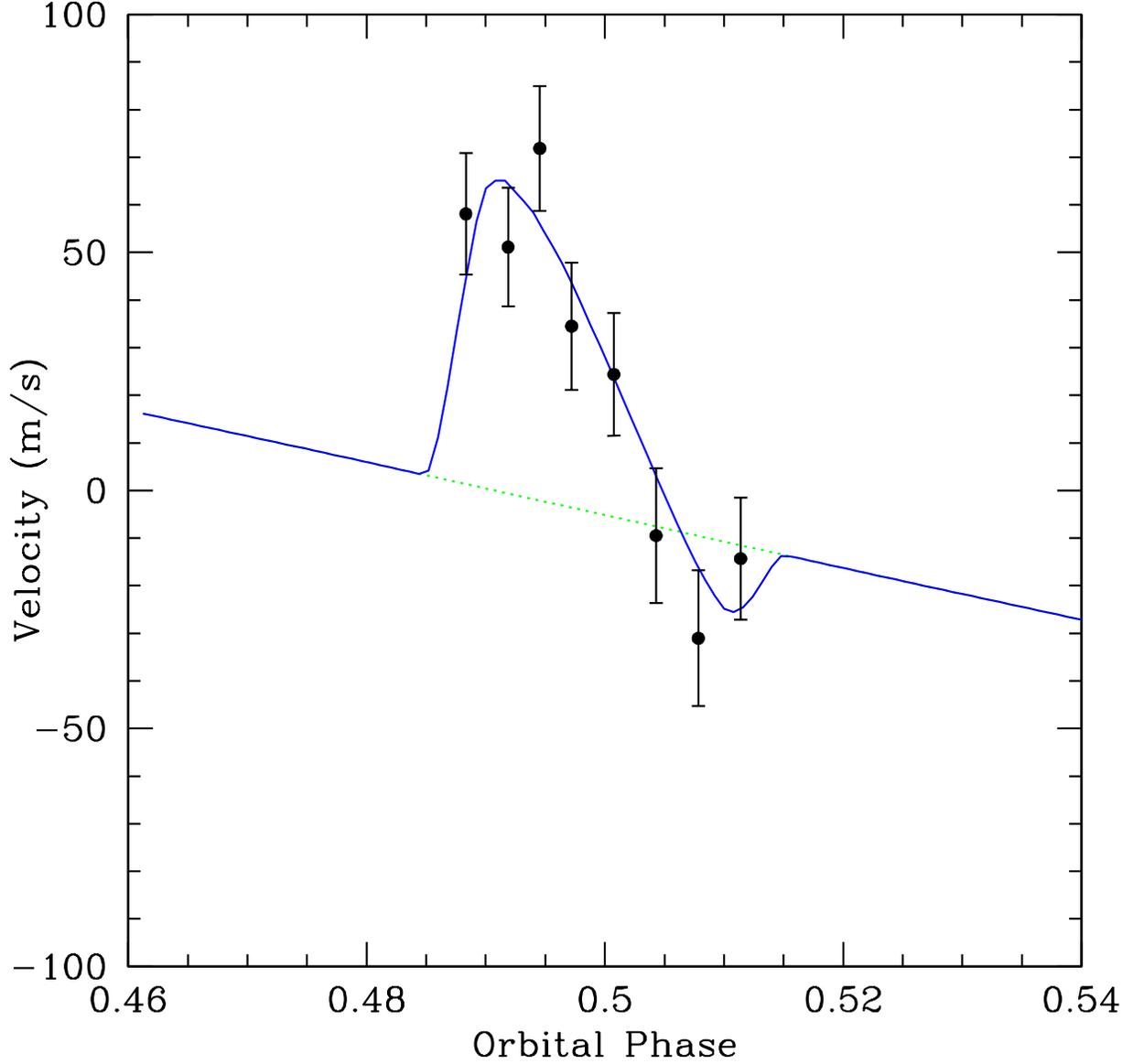}
\caption{
The Rossiter-McLaughlin effect for \koicur. The observed radial velocities
obtained with HIRES on the Keck 1 Telescope are plotted
together with the predicted velocities from the best-fit model of the
Rossiter-McLaughlin effect.  The high velocities near ingress and low
velocities near egress imply a prograde orbit.  The asymmetry
in the model curve indicates a projected tilt of the orbit relative to the
star's equator, as well as a transit with a large impact parameter
on the star.
\label{fig:rossiter}}
\end{figure}

We have carried out a wide array of observations of Kepler-8 that
pinpoint the existence and properties of the exoplanet.  The \emph{Kepler}
photometric measurements, with $\sim$0.1 mmag errors in 30 min
intervals, are tightly fit by a model of a transiting exoplanet
(Figure \ref{fig:lightcurve}).  This good fit to the inflections of the photometric data
provide immediate support for the planet interpretation, with few
plausible alternative interpretations except for a background, diluted
eclipsing binary system having just the right brightness and radius
ratio to mimic a planet.  We note that, unlike ground-based transit
work that has lower photometric precision, the \emph{Kepler} photometry is so
precise that blends of eclipsing binaries are more readily identified
from the photometry alone.   

Nonetheless, to test the unlikely possibility of a blend, we carried
out a battery of astrometric tests to hunt for an eclipsing binary.
The resulting steadiness of the position of the primary star, \koicur,
during and out of transit, argues against any eclipsing binary in the
photometric aperture.  We further carried out both adaptive optics
imaging and speckle interferometric measurments to hunt for faint
eclipsing binaries located within an arcsec of the \koicur .  None was
found, further diminishing the chance of such a masquerade.  We
followed with high resolution spectroscopy at both low and high
signal-to-noise ratio, finding no evidence of double lines nor rapid
rotation.  Further support for the planet interpretation came from the
precise radial velocities that varied in phase with, and had the same
period as, the photometric light curve, further supporting the planet
model and constraining the planet mass.

\begin{figure}
\plotone{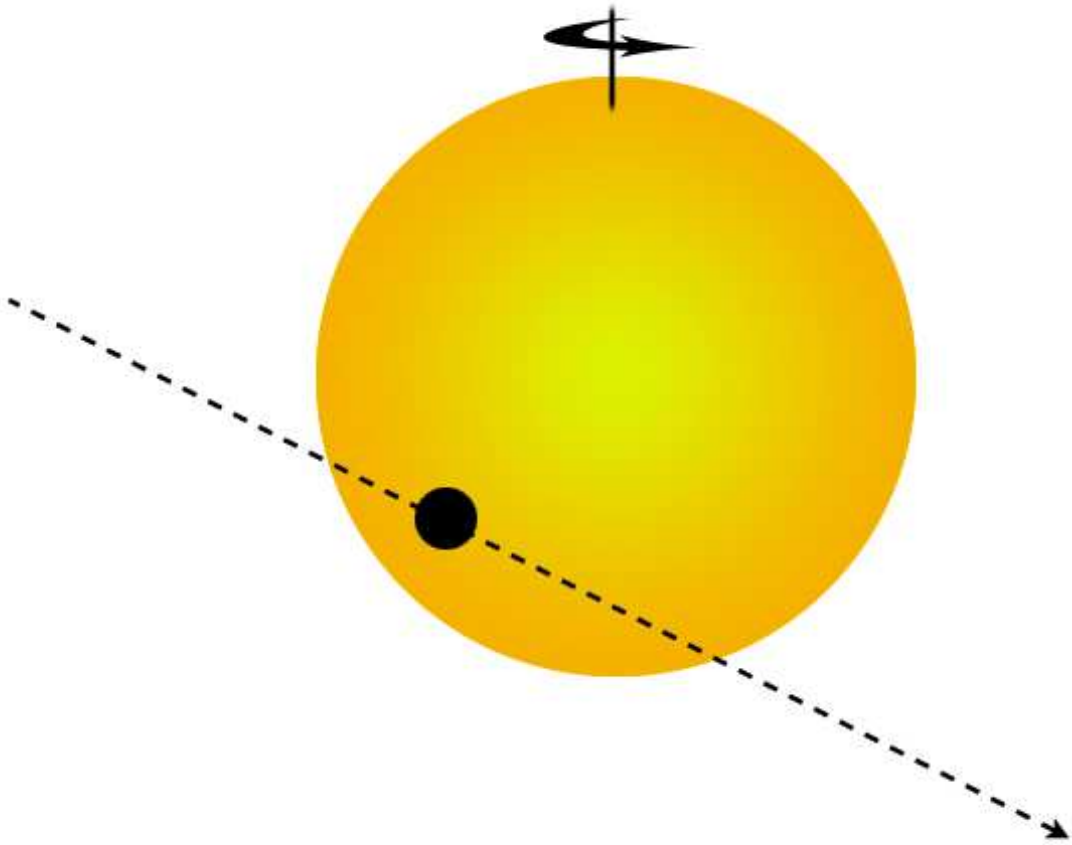}
\caption{ 
A sketch of the transit geometry for \koicur.  The \emph{Kepler}
  photometry and R-M measurements set the impact parameter, 
$b$ = \koicurLCimp, and the angle projected on the sky between the  
star's spin axis and the normal to the orbital plane,                                                          
$\lambda =  -27^{\circ}$, indicating a prograde orbit with a moderate inclination.    
\label{fig:transit_sketch}}
\end{figure}

Finally, the Rossiter-McLaughlin effect confirmed independently the
planet interpretation and provided further geometrical information
about the orbit, notably $\lambda = -26.9 \pm 4.6^{\circ}$.  
A sketch of the star and planet's orbit are shown in (Figure \ref{fig:transit_sketch}).
Remaining unknown is the inclination of the star's rotation axis, as
indicated in \ref{fig:transit_sketch}.  But
continued photometry during the \emph{Kepler} mission lifetime may
reveal a photometric periodicity caused by the rotation of spots
around the star.  The resulting rotation period, coupled with the
measured rotational \vsini = \koicurSMEvsini of the star, will allow
the star's inclination to be measured, putting the R-M geometry on
firmer ground.

The suite of ground-based observations obtained for \koicur described above
over-constrained a set of parameters that were obligated to mutually
agree, including such subtle issues as the limb-darkening, the stellar
rotation rate, and stellar radius, mass, and density.  Thus, the R-M
effect can, in general, be used to confirm the existence of a
planetary companion in the case where standard follow-up observing
procedures may provide somewhat ambiguous or equivocal results.
Examples would be \emph{Kepler} stars that are sufficiently saturated that
measurement of photocenter shifts during transits is problematic, or
stars like \koicur with large \vsini that make orbit determination via
precise RV measurement difficult.

Observations of the Rossiter-McLaughlin effect in dozens of transiting
planets, discovered from the ground and from \emph{Kepler}, offer an
excellent opportunity to determine the distribution of orbital geometries of the
short-period planetary systems in general.  These planetary
systems will continue to be studied in an extraordinarily unform and
consistent manner with the same set of tools.  The sample selection
effects are extremely well understood and documented.  The \emph{Kepler}
lightcurves are of unprecidented precision, allow sensitive searches
for additional planets in the system via transit timing variations,
and significantly tighter limits to be placed in the future on the
orbital eccentricity.  Coupling these data with R-M measurements of
the orbital alignment in these systems will allow us to correlate the
observed properties with the degree of orbital alignment, and thus to
search for the physical mechanisms causing observed misalignments.  We
will also have the information to be able to begin to back-out the
physical angle between the spin and orbital angular momentum vectors,
not just the projection of these angles on the plane of the sky.

\acknowledgements 
Funding for this Discovery mission is provided by
NASA's Science Mission Directorate.  Many people have contributed to
the success of the \emph{Kepler Mission}, and it is impossible to acknowledge
them all.   Valuable advice and assistance were provided
by 
Willie Torres, Riley Duren. M.~Crane, and D.~Ciardi.
Special Technical help was provided by Carly Chubak, G. Mandushev, and
Josh Winn.  We thank E.~Bachtel and his team at Ball Aerospace
for their work on the \emph{Kepler} photometer and R.~Thompson for
key contributions to engineering; and C.~Botosh, 
and J.~Fanson, for able management.  GWM thanks acknowledges support from
NASA Cooperative Agreement NNX06AH52G. 

{\it Facilities:} Kepler.



\begin{deluxetable}{lcc}
\tabletypesize{\scriptsize}
\tablewidth{0pc}
\tablecaption{System Parameters for \koicur \label{tab:parameters}}
\tablehead{\colhead{Parameter}	& 
\colhead{Value} 		& 
\colhead{Notes}}
\startdata
\sidehead{\em Transit and orbital parameters}
Orbital period $P$ (d)				& \koicurLCP		& A	\\
Midtransit time $E$ (HJD)			& \koicurLCT		& A	\\
Scaled semimajor axis $a/\rstar$		& \koicurLCar		& A	\\
Scaled planet radius \rpl/\rstar		& \koicurLCrprstar	& A	\\
Impact parameter $b \equiv a \cos{i}/\rstar$	& \koicurLCimp		& A	\\
Orbital inclination $i$ (deg)			& \koicurLCi 		& A	\\
Orbital semi-amplitude $K$ (\ms)		& \koicurRVK		& A,B	\\
Orbital eccentricity $e$			& 0 (adopted)		& A,B	\\
\sidehead{\em Observed stellar parameters}
Effective temperature \teff\ (K)		& \koicurSMEteff	& C 	\\
Spectroscopic gravity \logg\ (cgs)		& \koicurSMElogg	& C	\\
Metallicity \feh				& \koicurSMEfeh		& C	\\
Projected rotation Velocity \vsini\ (\kms)	& \koicurSMEvsini	& C	\\
Absolute (Helio) radial velocity (\kms)		& \koicurRVmean       & B \\
\sidehead{\em Derived stellar parameters}
Mass \mstar (\msun)				& \koicurYYmlong	& C,D	\\
Radius \rstar (\rsun)  				& \koicurYYrlong	& C,D	\\
Surface gravity \loggstar\ (cgs)		& \koicurYYlogg		& C,D	\\
Luminosity \lstar\ (\lsun)			& \koicurYYlum		& C,D	\\
Absolute V magnitude $M_V$ (mag)		& \koicurYYmv   	& D	\\
Age (Gyr)					& \koicurYYage		& C,D	\\
Distance (pc)					& \koicurXdist		& D	\\ 
\sidehead{\em Planetary parameters}
Mass \mpl\ (\mjup)				& \koicurPPm		& A,B,C,D	\\
Radius \rpl\ (\rjup)				& \koicurPPr		& A,BC,D	\\
Density \rhopl\ (\gcmc)				& \koicurPPrho		& A,B,C,D	\\
Surface gravity \loggpl\ (cgs)			& \koicurPPlogg		& A,B,C,D	\\
Orbital semimajor axis $a$ (AU)			& \koicurPParel		& E	\\
Equilibrium temperature \teq\ (K)		& \koicurPPteq		& F
\enddata
\tablecomments{\\
A: Based on the photometry.\\
B: Based on the radial velocities.\\
C: Based on an SME analysis of the Keck-HIRES spectra,\\
D: Based on the Yale-Yonsei stellar evolution tracks.\\
E: Based on Newton's version of Kepler's Third Law and total mass.\\
F: Assumes Bond albedo = 0.1 and complete redistribution.\\
}
\end{deluxetable}


\begin{deluxetable}{rrrrrr}
\tablewidth{0pc}
\tablecaption{Relative Doppler Velocity Measurements of \koicur{}\label{tab:velocities}}
\tablehead{
\colhead{HJD}                           &
\colhead{RV}                            &
\colhead{\ensuremath{\sigma_{\rm RV}}}  \\
\colhead{(-2450000)}                    &
\colhead{(\ms)}                         &
\colhead{(\ms)}                         &
}
\startdata
  4984.040 &  -22.15 &   34.5  \\
  4985.043 &   52.45 &   34.5  \\
  4986.065 &  -42.39 &   34.0  \\
  4987.060 &  -74.37 &   33.6  \\
  4988.027 &   63.04 &   33.7  \\
  4988.973 &   98.87 &   33.8  \\
  4995.103 &   46.58 &   34.2  \\
  5014.891 &  -64.61 &   33.7  \\
  5015.975 &  -29.97 &   33.4  \\
  5017.020 &  -28.28 &   33.3  \\
  5075.788 &   41.65 &   33.9  \\
  5109.850 & -122.00 &   32.8  \\
  5110.790 &  -67.42 &   31.9  \\
  5133.839 &   49.58 &   32.8  \\
  5133.718 &   58.11 &   32.6  \\
  5133.732 &   51.13 &   32.5  \\
  5133.747 &   71.85 &   32.7  \\
  5133.761 &   34.51 &   32.8  \\
  5133.776 &   24.38 &   32.6  \\
  5133.790 &   -9.49 &   33.2  \\
  5133.805 &  -31.04 &   33.2  \\
  5133.821 &  -14.32 &   32.6  \\
  5134.740 &  -23.48 &   31.9  \\
  5135.790 &   48.50 &   32.6  \\
\enddata
\end{deluxetable}


\end{document}